\documentclass[%
 superscriptaddress,
 amsmath,amssymb,
 aps,
twocolumn
]{revtex4-1}

\usepackage{graphicx}
\usepackage{dcolumn}
\usepackage{bm}

\newcommand{\argmax}{\mathop{\rm arg~max}\limits}

\begin{document}

\preprint{APS/123-QED}

\title{Efficient Construction Method for Phase Diagrams Using Uncertainty Sampling}

\author{Kei Terayama}
\email{kei.terayama@riken.jp}
\affiliation{%
 RIKEN Center for Advanced Intelligence Project, Tokyo 103-0027, Japan
}%
\affiliation{%
 Medical Sciences Innovation Hub Program, RIKEN Cluster for Science, Technology and Innovation Hub, Kanagawa 230-0045, Japan 
}%
\affiliation{%
Graduate school of Medicine, Kyoto University, Kyoto 606-8507, Japan 
}%

\author{Ryo Tamura}
\thanks{K. Terayama and R. Tamura contributed equally to this work.}
\affiliation{%
 International Center for Materials Nanoarchitectonics (WPI-MANA), 
National Institute for Materials Science, Ibaraki 305-0044, Japan
}%
\affiliation{%
 Research and Services Division of Materials Data and Integrated System, 
National Institute for Materials Science, Ibaraki 305-0047, Japan
}%
\affiliation{%
Graduate school of Frontier Sciences, the University of Tokyo, Chiba 277-8568, Japan
}%

\author{Yoshitaro Nose}
\affiliation{%
Department of Materials Science and Engineering, Kyoto University, Kyoto 606-8501, Japan
}%
\author{Hidenori Hiramatsu}
\affiliation{%
Laboratory for Materials and Structures, Institute of Innovative Research, Tokyo Institute of Technology, Yokohama 226-8503, Japan
}%
\affiliation{%
Materials Research Center for Element Strategy, Tokyo Institute of Technology, Yokohama 226-8503, Japan
}%

\author{Hideo Hosono}
\affiliation{%
Laboratory for Materials and Structures, Institute of Innovative Research, Tokyo Institute of Technology, Yokohama 226-8503, Japan
}%
\affiliation{%
Materials Research Center for Element Strategy, Tokyo Institute of Technology, Yokohama 226-8503, Japan
}%

\author{Yasushi Okuno}
\affiliation{%
 Graduate school of Medicine, Kyoto University, Kyoto 606-8507, Japan 
}%

\author{Koji Tsuda}
\email{tsuda@k.u-tokyo.ac.jp}
\affiliation{%
Graduate school of Frontier Sciences, the University of Tokyo, Chiba 277-8568, Japan
}%
\affiliation{%
 Research and Services Division of Materials Data and Integrated System, 
National Institute for Materials Science, Ibaraki 305-0047, Japan
}%
\affiliation{%
 RIKEN Center for Advanced Intelligence Project, Tokyo 103-0027, Japan
}%

\date{\today}

\begin{abstract}
We develop a method to efficiently construct phase diagrams using machine learning.
Uncertainty sampling (US) in active learning is utilized to intensively sample around phase boundaries.
Here, we demonstrate constructions of three known experimental phase diagrams by the US approach.
Compared with random sampling, the US approach decreases the number of sampling points to about 20\%.
In particular, the reduction rate is pronounced in more complicated phase diagrams.
Furthermore, we show that using the US approach, undetected new phase can be rapidly found, and smaller number of initial sampling points are sufficient. 
Thus, we conclude that the US approach is useful to construct complicated phase diagrams from scratch and will be an essential tool in materials science.
\end{abstract}

\pacs{Valid PACS appear here}
\maketitle


\section{\label{sec:intro}Introduction}

Phase diagrams are crucial in materials development because they contain extremely useful information. However, numerous syntheses and measurements are necessary to complete a phase diagram. Thus, this indispensable task occupies a large percentage of materials discovery.

In combinatorial materials science,
machine learning techniques have been applied to construct phase diagrams\cite{Long-2007,Kusne-2014,Bunn-2015,Iwasaki-2017,Xue-2017,Pradhan-2018}.
In this field, a large amount of materials in the phase diagram can be obtained simultaneously by high-throughput materials synthesis. Then, from measurement results such as XRD patterns, categories of many synthesized materials should be rapidly determined to complete phase diagrams. To realize automatically categorization, clustering and matrix factorization are utilized.

On the other hand, recent materials informatics studies aim to develop novel materials with the smallest number of syntheses or first-principles calculations as possible with aid of machine learning. 
Many successful examples have been reported using both experiments and simulations\cite{Pilania-2013,Meredig-2014,Seko-2015,Ueno-2016,Ikebata-2017,Ju-2017,Kim-2017,Pilaniay-2017,Bombarelli-2018,Sawada-2018,Sumita-2018}. 
In these investigations, machine learning efficiently recommends a candidate material possessing the desired properties even if a limited materials data is existed.
In accordance with this idea, 
we aim to propose some materials which should be synthesized to complete phase diagrams by machine learning.
If an appropriate proposal is realized, a reliable phase diagram can be obtained, even if the number of synthesized materials is small. This problem setting resembles that in active learning.

Active learning is a learning framework that sequentially selects an informative sample to classify and checks its label in order to maximize the classification accuracy with fewer labeled data. Thus, we speculate that active learning is an essential tool to efficiently construct a phase diagram. A previous study employed an active learning method that uses a Gaussian process to sample the phase diagram \cite{dai2018efficient}. Although this method dramatically reduces the number of sampling points, the demonstration was only performed using a phase diagram with only two kinds of phases. Furthermore, this method would be difficult to apply in cases where the multiple phases exist. To improve practicability, herein we propose a different active learning method that uses uncertainty sampling (US) \cite{settles2012active} to efficiently construct a phase diagram.

US is a methodology that selects a sampling point with the most uncertainty as calculated by a machine learning-based classification model as an informative sample.
The most uncertain data is typically located near classification boundaries (phase boundaries). 
This US approach can be applied to any numbers of parameters (dimensions in a phase diagram) and categories (kinds of phases).

In this paper, the US approach is used to construct phase diagrams. This study reveals the following:
(1) Phase boundaries can be efficiently obtained and accurate phase diagram can be drawn even if the number of sampling points is small.
(2) Because undetected new phases in a phase diagram can be rapidly found, this approach is more useful to construct complicated phase diagrams.
(3) Fewer initial sampling points are sufficient, making the US approach well suited to construct phase diagrams from scratch.
These facts strongly suggest that the US approach will be a powerful tool to construct phase diagrams in materials science.
Our implementation is available on GitHub at https://github.com/tsudalab/PDC/.

The rest of the paper is organized as follows. Section II introduces details of our method based on US to efficiently construct phase diagrams. To estimate the probability of phases at each point from already checked points, the label propagation and label spreading methods are adopted. Furthermore, the evaluation methods of uncertainty, that is, least confident, margin sampling, and entropy-based approach, are explained. In Sec. III, the US approach is used to construct three known phase diagrams: H$_2$O under lower and higher pressures, and a ternary phase diagram of glass-ceramic glazes. 
The US approach can efficiently sample to complete a complicated phase diagram from scratch. Section IV addresses the case with experimental constraints.
The results with and without imposed constraints are comparable.
Section V is the discussion and summary.

\section{\label{sec:method}Method based on Uncertainty Sampling}

\begin{figure*}
\centering 
  \includegraphics[clip,width=0.8\linewidth]{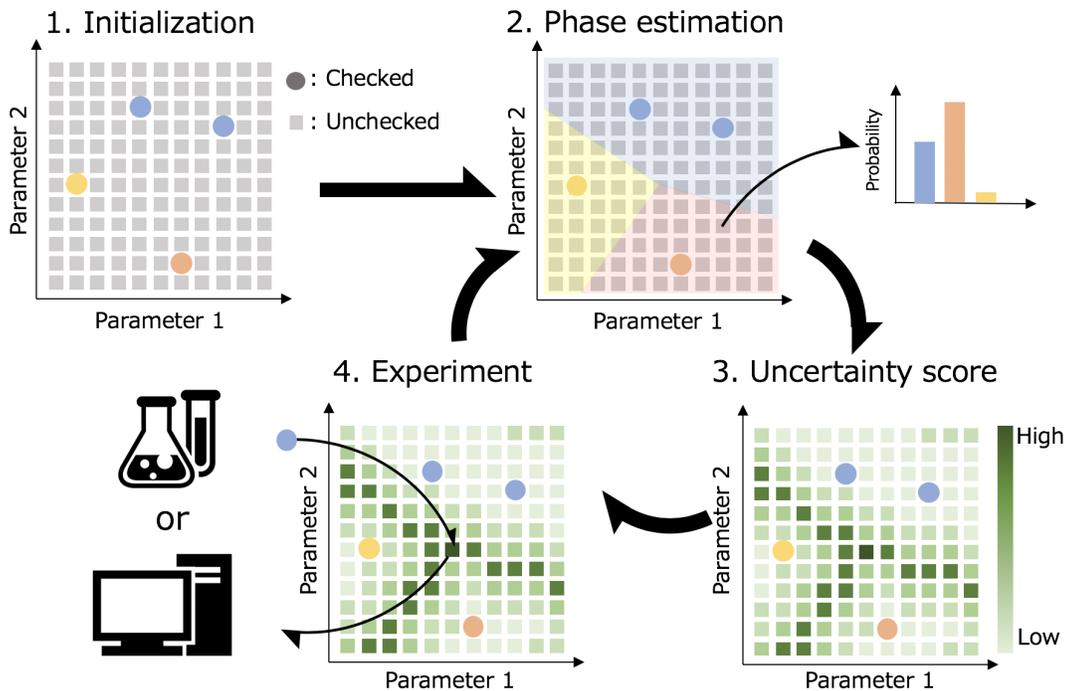}
  \caption{Overview of efficient phase diagram construction based on the uncertainty sampling approach.}
\label{fig:overview}
\end{figure*}

This section presents the framework for phase diagram construction using US.
Figure~\ref{fig:overview} overviews our procedure.
First, several points are selected, and their phases are determined by experiments or simulations (1. Initialization). Next, the probability distributions of the phases are calculated for all the points in the parameter space using a machine learning technique (2. Phase estimation). From the probability distributions, the uncertainty scores are calculated for the all unchecked points in the parameter space (3. Uncertainty score). Afterwards, an experiment or a simulation of the point with the highest uncertainty score is performed (4. Experiment).
Steps 2-4 are repeated to construct an accurate phase diagram with a smaller number of sampling points.
Below each step is described in detail.

\subsection{Initialization}
The regions of parameter space and parameter candidates are prepared in advance. Parameter space can have two or more dimensions. As the initialization step, several points are selected, and their phases are determined by experiments or simulations. 
The points can be selected randomly or manually.
This paper adopts random selection.

\subsection{Phase estimation}
The probabilities of the observed phases are estimated for all unchecked points. 
This probability distribution is written as $P(p | \mathbf{x})$, where $\mathbf{x}$ is the position vector of each unchecked point
and $p$ is the label of phases, which are already observed.
From this distribution, an estimated phase diagram is drawn by choosing the phase with the highest probability, e.g. $\argmax_p P(p|\mathbf{x})$. 
Herein we adopt two representative estimation methods of probabilities: label propagation (LP) \cite{zhu2003semi} and label spreading (LS) \cite{zhou2004learning}. 
These are kinds of the semi-supervised learning, which makes use of not only labeled but also unlabeled data for learning. 
In these methods, the probability of each point is calculated by propagating the label information to nearby points. 
The probability of a phase $p$ at $\mathbf{x}$ calculated by LP equals to one to reach the phase $p$ first by random walk from $\mathbf{x}$.
In LP, the labels of the checked points are fixed.
On the other hand, in LS, the labels of checked points can be changed depending on the surrounding circumstances.
Thus, LS is effective when the label noise is large.

\subsection{Uncertainty score}
The uncertainty score defined as $u(\mathbf{x})$ is calculated to determine the next candidate in a phase diagram from the estimation result of the probability distributions $P(p | \mathbf{x})$.
In this paper, we adopt three representative methods of the US strategy: Least Confident (LC) \cite{lewis1994sequential}, Margin Sampling (MS) \cite{scheffer2001active}, and Entropy-based Approach (EA) \cite{shannon1948mathematical}. For each point $\mathbf{x}$, the LC $u_{\mathrm{LC}} (\mathbf{x})$, MS $u_{\mathrm{MS}} (\mathbf{x})$, and EA $u_{\mathrm{EA}} (\mathbf{x})$ scores are calculated as follows:
\begin{eqnarray}
  u_{\mathrm{LC}} (\mathbf{x}) &=& 1 - \max_{p} P(p|\mathbf{x}), \\
  u_{\mathrm{MS}} (\mathbf{x}) &=& 1 - [P(p_1|\mathbf{x})-P(p_2|\mathbf{x})],\\
  u_{\mathrm{EA}} (\mathbf{x}) &=& -\sum_{p}P(p|\mathbf{x})\log{P(p|\mathbf{x})},
\end{eqnarray}
where $P(p_1|\mathbf{x})$ and $P(p_2|\mathbf{x})$ in $u_{\mathrm{MS}} (\mathbf{x})$ mean the highest and second highest probabilities at $\mathbf{x}$. 
From the definitions, the uncertainty scores become higher when the probabilities of each phase are all the same. 
The LC score is only influenced by the highest probability at each point, while the MS score is affected by the first and second highest probabilities. For the EA score, the whole distribution is taken into account.

The next candidate is determined from the unchecked points with the highest uncertainty score, e.g. $\argmax_{\mathbf{x}} u(\mathbf{x})$.
Then an experiment or a simulation is performed for this point.
If an undetected phase is obtained, the next step performs a phase estimation that includes the new phase.
To handle data uniformly, the parameters are normalized using the min-max normalization \cite{jain2005score} for a phase estimation and an evaluation of uncertainty score.

\section{\label{sec:demo} Phase Diagram Construction by Uncertainty Sampling}

We report the performances of the proposed strategies based on US compared to random sampling (RS) for three known phase diagrams: H$_2$O under lower pressure (H$_2$O-L), H$_2$O under higher pressure (H$_2$O-H)\cite{Bridgman-1912,Mercury-2001,ice_h}, and the ternary phase diagram of glass-ceramic glazes of SiO$_2$, Al$_2$O$_3$, and MgO (SiO$_2$-Al$_2$O$_3$-MgO) \cite{lesniak2016microstructure}.
Figures~\ref{fig:PD_example} (a), (b), and (c) show these experimental phase diagrams.
Here, the next point in RS is randomly selected from the unchecked points and the phase diagram is estimated using the phase estimation methods described above. 
However, the information of the estimated phase diagram is not used to select the next point in RS.

\subsection{Sampling results} \label{sec:sampling}

\begin{figure*}
\centering 
  \includegraphics[clip,width=0.8\linewidth]{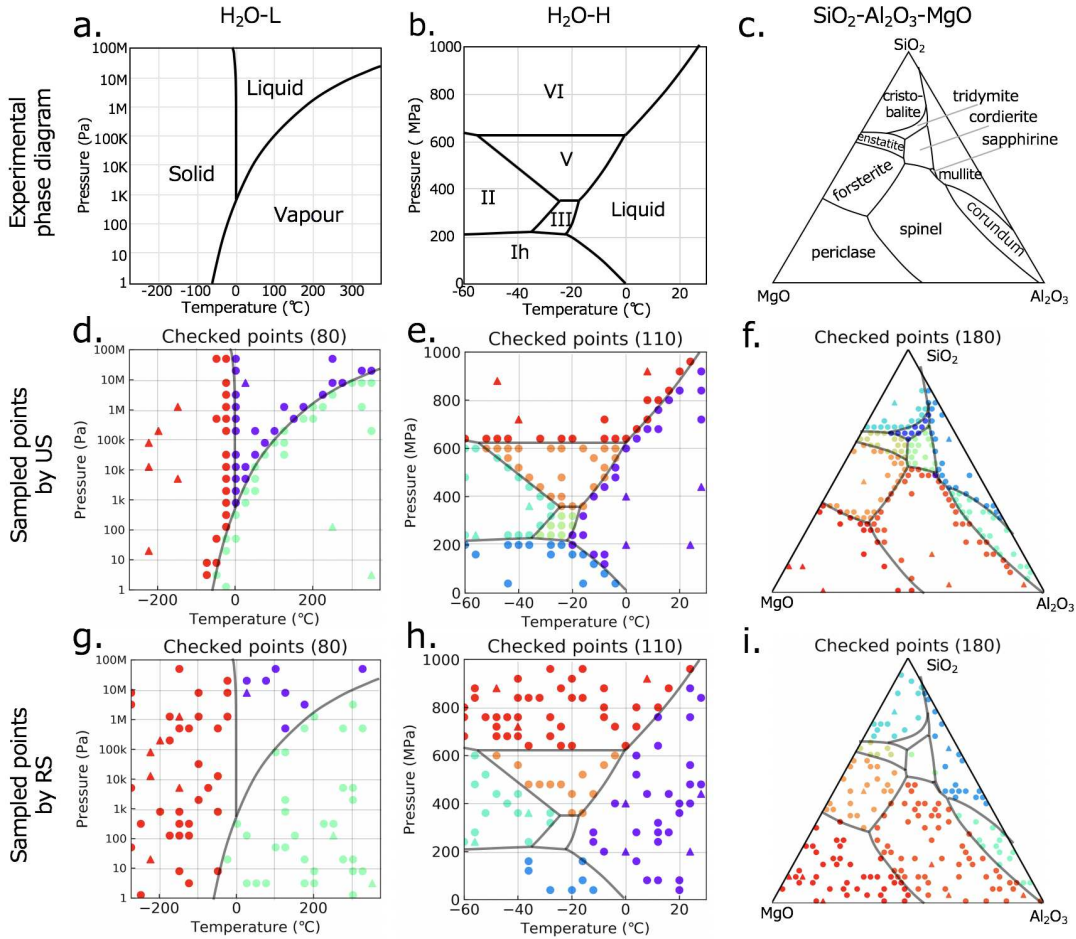}
  \caption{Phase diagrams of H$_2$O-L (a), H$_2$O-H (b), and SiO$_2$-Al$_2$O$_3$-MgO (c) and examples of samplings by using the US and RS approaches. (a), (b), and (c) have 3, 6, and 10 phases, respectively. (d), (e), and (f) show nine initial points randomly selected (triangles) and sampled points (circles) by the LP+LC method. Numbers of sampling points are 80, 110, and 180. (g), (h), and (i) show sampling points by the RS method. Numbers of sampling points in (g), (h), and (i) are the same as (d), (e), and (f). 
 Relatively small phases such as ice III in (h) and tridymite and sapphirine in (i) are not found by the RS method.
  }
\label{fig:PD_example}
\end{figure*}

Figures~\ref{fig:PD_example} (d), (e), and (f) show examples of the sampled points by the US approach.
In these demonstrations, LP is utilized as the phase estimation method and LC (LP+LC) is used to evaluate the uncertainty score.
For each case, nine points are randomly selected as the initialization.
In total, the sampled points are 80, 110, and 180 for H$_2$O-L, H$_2$O-H, and SiO$_2$-Al$_2$O$_3$-MgO, respectively. Triangles denote the initial points and circles are the sampled points by the US approach. The sampled points are distributed around the phase boundaries. 
On the other hand, Figs.~\ref{fig:PD_example} (g), (h), and (i) show the sampling results by RS. 
The nine triangles denote the initial points, which are located at the same positions as the US approach.
Since RS selects a number of points in regions away from the phase boundaries,
efficient sampling is not realized.
In addition, relatively small areas such as ice III in H$_2$O-H and tridymite
and sapphirine in SiO$_2$-Al$_2$O$_3$-MgO are difficult to find by RS.
However, these phases can be detected by the US approach, as shown in Figs. \ref{fig:PD_example} (e) and (f). These results suggest that the US approach can efficiently sample near the phase boundaries, allowing smaller phases to be rapidly detected.
As supplemental material,
movies of sampling behaviors for each case by the LP+LC approach compared with LP+RS are prepared (see Supplemental Movie 1).

\subsection{Quantitative comparison between uncertainty and random samplings}

We quantitatively compare the US approach with the RS approach.
To evaluate the quantitative accuracy of the estimated phase diagram by the LP or LS method, we adopted the macro average score based on the F1-score (Macro-F1), which is commonly used as an evaluation metric for classification problems in the machine learning community. 
This value denotes the difference between the experimentally obtained phase diagram and the estimated phase diagram.
The F1 score for a phase indexed by $p$ is the harmonic mean of precision $P(p)$ and recall $R(p)$, which is given as
\begin{eqnarray}
 F_1(p) = \frac{2 P(p) R(p)}{P(p)+R(p)},
\end{eqnarray}
where precision $P(p)$ is the number of points correctly estimated as $p$ (true positives) divided by the total number of points estimated as $p$. 
On the other hand, recall $R(p)$ is the number of true positives divided by the total number of true $p$ points.
A phase that has yet to be detected has an F1-score of 0. We calculated the Macro-F1 score by averaging the F1-scores of all the true phases. 
Thus, when the value of the Macro-F1 score is small value ($\ll 1$), the difference between the true and the estimated phase diagrams is very large.
A Macro-F1 score of 1 indicates that the estimated phase diagram exactly reproduces the true one.

\begin{figure*}
\centering 
  \includegraphics[clip,width=0.9\linewidth]{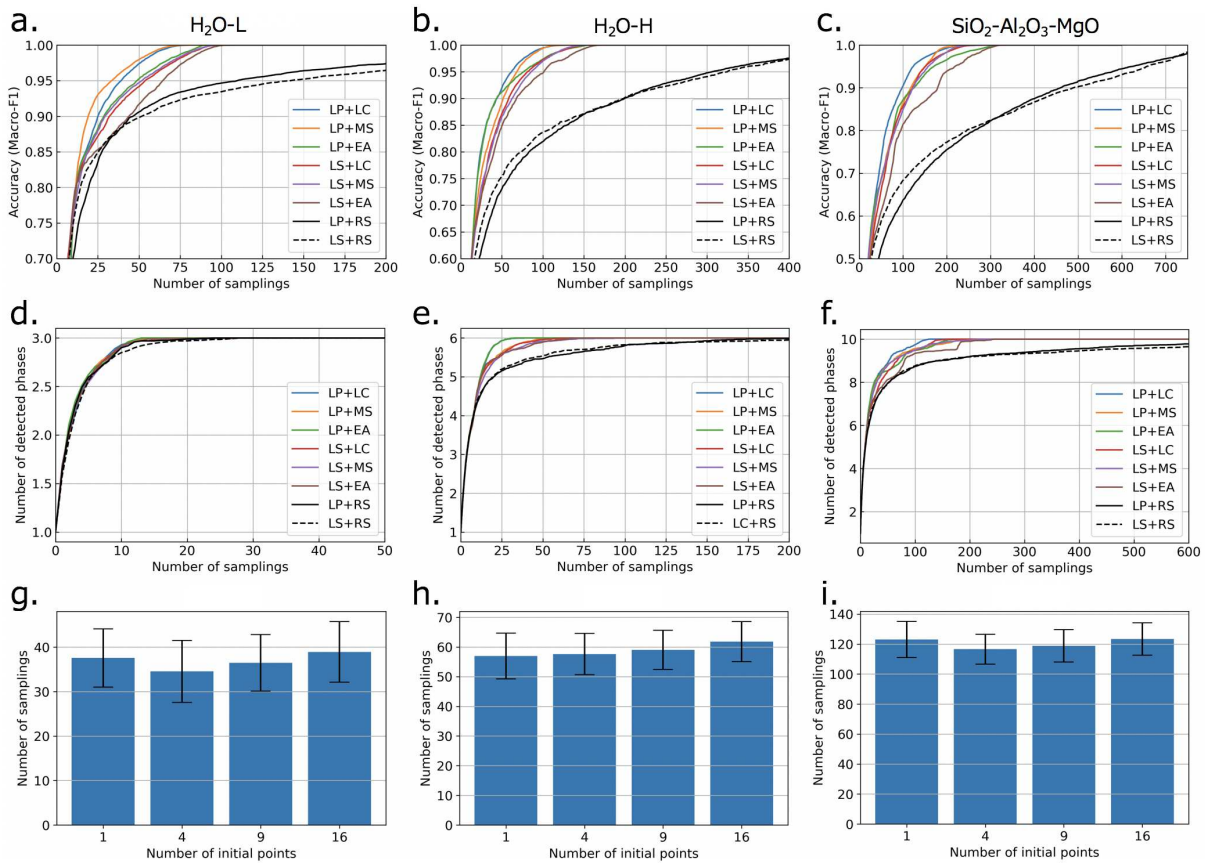}
  \caption{Performances of phase diagram construction using the US approach. Top row shows the accuracies (Macro-F1) of phase diagram construction as functions of the number of sampling points for H$_2$O-L (a), H$_2$O-H (b), and SiO$_2$-Al$_2$O$_3$-MgO (c).
  Colored lines are the results from the US approach. Black solid and dashed lines are from RS.
  Middle row shows the performances of phase detection. (d), (e), and (f) show the number of sampling points necessary to detect all phases in the phase diagrams of  H$_2$O-L (3 phases), H$_2$O-H (6 phases), and SiO$_2$-Al$_2$O$_3$-MgO (10 phases). 
  Bottom row shows the effects of the initial sampling points. (g), (h), and (i) show the numbers of sampling points to reach Macro-F1 of 0.95 by using LP+LC as functions of initial sampling points (1, 4, 9, and 16). 
  }
\label{fig:result}
\end{figure*}

The top row in Fig.~\ref{fig:result} shows the results of the Macro-F1 scores as functions of the number of sampling points for H$_2$O-L, H$_2$O-H, and SiO$_2$-Al$_2$O$_3$-MgO. Here, the initial sampling number is fixed to nine. Since these results depend on the selection of initial points, we repeated the trials 200 times using different initial points and averaged the results. The black lines (solid and dashed) depict the results of RS, and the other lines show the results by the US approach. The label of A+B means the combination of A as the phase estimation method (i.e., LP or LS) and B as the sampling method (i.e., LC, MS, EA, or RS). The combinations of LP+LC and LP+MS show relatively good performances compared with the other methods for the three phase diagrams. 
Compared with RS, any US approach can provide better Macro-F1 scores even if the number of sampling points is small.
Table \ref{tbl:result} summarizes the numbers of sampling points necessary to reach a Macro-F1 of 0.95.
From the viewpoint of the average from three phase diagrams, the number of sampling points could be reduced by 0.36, 0.20, and 0.20 for the three phase diagrams using the LP+LC method instead of the LP+RS method.
This result implies that for complicated phase diagrams, the US approach is more useful to produce it quickly.

For these phase diagrams, 
the Macro-F1 results indicate that the LP method is better suited than LS.
This demonstration employs cases where the phase boundaries are properly determined
and the outliers do not appear.
Since it is not necessary to consider the noise for the labels, LS does not work effectively.
Furthermore, we found that EA is not useful.
If the number of phases is small, MS is better suited, whereas LC is powerful when many phases exist in a phase diagram.
Thus, an efficient selection can be realized using LC to construct complicated phase diagrams.

\begin{table*}[]
\caption{\label{tbl:result}
Number of sampling points to reach an accuracy (Macro-F1) of 0.95 using the US approach and the RS approach. 
Parentheses denote the reduction rates of the US approach compared with the RS approach.
For the LP-based US approach, the numbers of sampling points of LP+LC, LP+MS, and LP+EA are divided by those of LP+RS,
while LS-based US approaches (LS+LC, LS+MS, and LS+EA) are divided by LS+RS.
Bolded values indicate the highest accuracy.}
\begin{ruledtabular}
\begin{tabular}{c|cccccc|cc}
 System & LP+LC & LP+MS & LP+EA & LS+LC & LS+MS & LS+EA & LP+RS & LS+RS \\
\hline
H$_2$O-L& 40 (0.36) & \textbf{34 (0.31)} & 49 (0.45) & 54 (0.38) & 52 (0.37) & 65 (0.46) & 110 & 142 \\
H$_2$O-H& \textbf{62 (0.20)} & 69 (0.22) & 77 (0.25) & 81 (0.25) & 86 (0.26) & 102 (0.31) & 307 & 325  \\
SiO$_2$-Al$_2$O$_3$-MgO& \textbf{124 (0.20)} & 151 (0.24)& 171 (0.27) & 151 (0.23) & 161 (0.24) & 224 (0.33) & 625 & 657 \\\hline
Average& \textbf{75.3 (0.25)} & 84.6 (0.26) & 99.0 (0.32) & 95.3 (0.29) & 98.7 (0.29) & 129 (0.37) & 347 & 375 \\
\end{tabular}
\end{ruledtabular}
\end{table*}

\subsection{Capability of New Phase Detection}

In Sec.~\ref{sec:sampling}, we showed that small phases are detected more quickly by using the US approaches than RS. In this subsection, we demonstrate how many sampling points are needed to detect all the phases in each phase diagram of H$_2$O-L, H$_2$O-H, and SiO$_2$-Al$_2$O$_3$-MgO. The middle row in Fig.~\ref{fig:result} shows the sampling number dependence of the numbers of detected phases which are averaged 200 independent runs.
Since H$_2$O-L has three large phases, the detection performances of the US approaches and RS are almost the same. In the case of H$_2$O-H, which has one small phase (ice III), all the phases are detected by LP+LC using 30 sampling points at most,
whereas RS requires more than 200 sampling points.
For SiO$_2$-Al$_2$O$_3$-MgO, over 600 sampling points are needed to find all the phases using RS because there are multiple small phases such as sapphirine, tridymite, and enstatite.
These results indicate that a small phase connected to boundaries of other large phases can be found relatively early using the US approach due to preferential investigation of areas near boundaries. 
Thus, we conclude that our US approach is a powerful tool to detect new phases in complicated phase diagrams.

\subsection{Effect of initial sampling}

We discuss the dependency of the initial sampling.
The bottom row in Fig.~\ref{fig:result} shows the average number of sampling points to reach Macro-F1 of 0.95 using the LP+LC approach as a function of the number of initial sampling.
To evaluate the average, 200 independent demonstrations are performed for different initial points.
Interestingly, the accuracy remains almost the same even as the number of initial sampling points increases.
Furthermore, in these cases, the optimum value of the initial sampling points is 1 or 4, and then at initial sampling, some phases are not detected.
This finding indicates that it is better to use information of a phase estimation by machine learning than blindly selecting points from the earlier stage to construct a phase diagram.
Consequently, the US approach is particularly useful when constructing new phase diagrams from scratch.

\section{\label{sec:const}Sampling with parameter constraint}

In the US approach described above, there are no restrictions on the change in parameters to select the next point.
However, there is often a huge cost to change all parameters in an experiment.
To address this problem, we construct a sampling method called Uncertainty Sampling with Parameter Constraint (USPC).
USPC constrains the changes in parameters.
To select the next point, candidate points are chosen under the condition where only one parameter is changed from the previous point. 
That is, for example in H$_2$O phase diagrams, candidate points are prepared along parallel or vertical directions from the previous point.
Then the one with the highest uncertainty score among the candidates is selected.  
The other steps are the same as the US approach. 
Note that if there are no candidates satisfying the condition, the next point is selected after removing the constraint. 
Thus, USPC can reduce the cost associated with a parameter change in the experiment.

\begin{figure*}
\centering 
  \includegraphics[clip,width=0.95\linewidth]{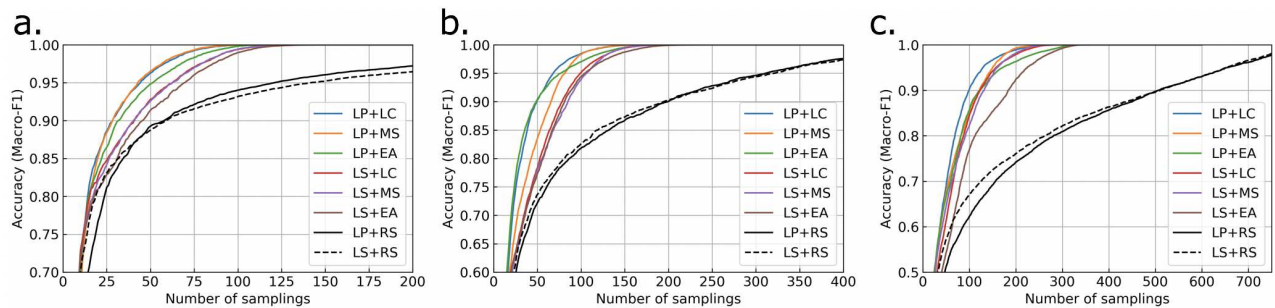}
  \caption{Performances of the USPC approach. 
  Colored lines are the results from the USPC approach. Black solid and dashed lines are from RSPC.
  }
\label{fig:result_const}
\end{figure*}

Figures~\ref{fig:result_const} (a), (b), and (c) show the results of the Macro-F1 scores as functions of the numbers of sampling points for H$_2$O-L, H$_2$O-H, and SiO$_2$-Al$_2$O$_3$-MgO by using the USPC approach, respectively.
Additionally, the results by RS with Parameter Constraint (RSPC) are plotted where the next point is randomly selected by imposing the same constraint as USPC.
USPC-based methods such as LP+LC and LP+MS show higher performances compared with RSPC, although the performances are somewhat lower than the results without the constraint (see Figs.~\ref{fig:result} (a), (b), and (c)). 
Furthermore, the numbers of sampling points to reach Macro-F1 of 0.95 are also successfully reduced to 0.37, 0.22, and 0.20 compared to RSPC (see Supplemental Table~S1).
The USPC approach shows similar tendencies for new phase detection and the effects on the initial sampling (see Supplemental Fig.~S1).
These results show that phase diagrams can be efficiently constructed even under the constraint suited for an experiment.
As supplemental material,
movies of the sampling behaviors for each cases by the LP+LC approach with parameter constraint compared with LP+RS are prepared (see Supplemental Movie 2).

\section{\label{sec:discussion}Discussion and Summary}
We proposed an efficient method to construct phase diagrams using uncertainty sampling (US). This method employs the next point with most uncertainty in a phase diagram assisted by machine learning. In general, the next point selected by this approach is located near phase boundaries, allowing the true phase boundary to be rapidly drawn. In our method, the uncertainty is evaluated using the probabilities of the observed phases at each point, which are obtained by the label propagation (LP) or label spreading (LS) methods.

By comparing the US approach with the random sampling, we confirmed that our approach can decrease the number of sampling points to 20 \% and still construct an accurate phase diagram. Furthermore, the US approach can find undetected new phase rapidly and smaller number of initial sampling points are sufficient to obtain an accurate phase diagram.
These advantages indicate that our method can make significant contributions, especially when deriving new complicated phase diagrams from scratch. 

We also considered the case where only one parameter is changed from the previous point when selecting the next candidate point, which is fitted to the conventional experimental setting. Even if such a constraint is imposed, this approach can realize efficient sampling to complete a phase diagram.
To strengthen the usability of our method,
we should construct new experimental phase diagrams using the US approach.
In this case, 
depending on the accuracy of the experiments to detect a phase,
LS might be more useful than LP to evaluate the uncertainty score due to the existence of noise.
These facts will be reported elsewhere.

From a different perspective, the US approach provides useful information about the reliability of the experiments when the phase at each point is determined. For all points in a phase diagram, the probabilities of each phase are evaluated in our approach. Thus, if the probability of the detected phase by experiments is extremely small, it may be an indicator that the experiment is wrong. This would be important information to construct valid phase diagrams.

The US approach can realize efficient sampling for phase diagrams. Therefore, we believe that the US approach will accelerate the speed to discover new materials. Hence, this method will become an essential tool in materials science.

\begin{acknowledgments}
We thank Fumiyasu Oba and Masato Sumita for the useful discussions.
This article is based on the results obtained from a project subsidized by 
the New Energy and Industrial Technology Development Organization (NEDO),
the ``Materials Research by Information Integration'' Initiative (MI2I) project, 
and Core Research for Evolutional Science and Technology (CREST) [Grants JPMJCR1502 and JPMJCR17J2] from the Japan Science and Technology Agency (JST). 
This work was also supported by the Ministry of Education, Culture, Sports, Science, and Technology of Japan (MEXT) as a ``Priority Issue on Post-K Computer'' (Building Innovative Drug Discovery Infrastructure through Functional Control of Biomolecular Systems).
K. Ts. and H. Hi. were supported by a Grant-in-Aid for Scientific Research on Innovative Areas ``Nano Informatics'' [Grant 25106005] from the Japan Society for the Promotion of Science (JSPS) and Support for Tokyotech Advanced Research (STAR), respectively.
\end{acknowledgments}

\providecommand{\noopsort}[1]{}\providecommand{\singleletter}[1]{#1}%

\end{document}